\documentstyle[aps,epsf,palatino,amstex]{revtex}
\begin{document}

\draft

\title{Quantum Games}
\author{Jens Eisert and Martin Wilkens}
\address{
Institut f{\"u}r Physik, Universit{\"a}t Potsdam,
14469 Potsdam,
Germany}
\date{\today}
\maketitle
\begin{abstract}
In these lecture notes we investigate the implications
of the identification of strategies with
quantum operations in game theory
beyond the results presented in
[J.\ Eisert, M.\ Wilkens, and M.\ Lewenstein,
        Phys.\ Rev.\ Lett.\ {\bf 83}, 3077 (1999)].
After introducing a general 
framework, we study quantum games
with a classical analogue in order to flesh out
the peculiarities of game theoretical settings in the 
quantum domain.
Special emphasis is given to
a detailed investigation of 
different sets of quantum strategies.
\end{abstract}
\pacs{PACS-numbers: 03.67.-a, 03.65.Bz, 02.50.Le}
\section{Introduction}

Game theory is the theory of decision making, which
provides powerful 
tools
for investigating situations in which several parties
make decisions 
%
according to their
personal interest \cite{NeuMor47,Mye91,Pou92,BooksOnGames}. 
It 
gives an account of how 
the parties would decide in a situation 
which involves contest, rivalry, or struggle.
Such have been found to be relevant to
social sciences, biology, or economics.
Of particular interest to the theory
are games of incomplete information in which  
the parties may choose their plan of action 
with complete knowledge of the situation on rational 
grounds, but without knowing what decision the other parties
have actually taken.

\begin{figure}
\centerline{
	\epsfxsize=5.0cm
       \epsfbox{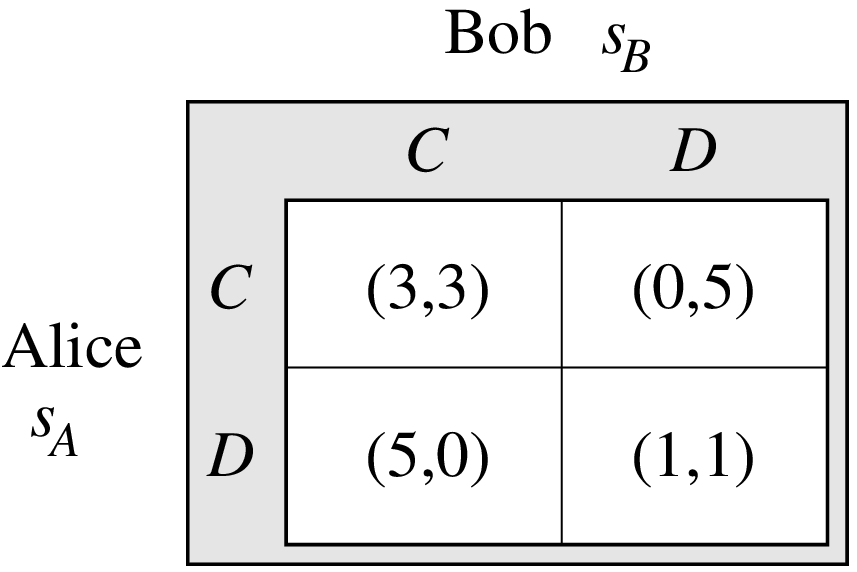}
}
\medskip

\caption{The pay-off matrix in the Prisoners' Dilemma game.
The first entry refers to Alice's pay-off, the second
to Bob's. If both players
cooperate, they both get $3$ units pay-off. 
If Bob defects and Alice happens
to cooperate, he obtains $5$ units, while Alice 
is in the unfortunate situation in which she does
not receive any pay-off at all. Bob faces the same
situation if he chooses to cooperate while 
Alice prefers to defect. If both defect, they 
get only $1$ unit pay-off.}
\label{Prison}
\end{figure}

One important two player 
game is the so-called {\it Prisoners' Dilemma}\/ \cite{Prisoner}.
In this game
two players
-- 
in the following referred to as Alice and Bob -- can 
independently decide whether they intend to ''cooperate'' 
or ''defect''. 
Being well aware of the 
consequences of their decisions 
the players obtain a certain pay-off
according to their respective choices.
This pay-off provides 
a quantitative characterisation of 
their personal preferences.
Both players are assumed to want to maximise
their individual pay-off, yet
they must pick their choice without
knowing the other player's decision.
Fig.\ \ref{Prison}
indicates the pay-off of Alice and Bob.
The players face a dilemma since
rational reasoning
in such a situation 
dictates the players to defect, although they
would both benefit from mutual cooperation.
%
As Alice is better off with defection regardless
of Bob's choice, she will defect. The game being symmetric,
the same argument applies to Bob.

Formally, Alice has two basic choices, meaning that she
can select from two possible 
{\it strategies}\/ $s_A=C$
(cooperation) and $s_A=D$ (defection). Bob
may also take $s_B=C$ or $s_B=D$. 
The game is defined by these possible strategies
on the one hand, and on the other hand 
by a specification of how to evaluate the 
pay-off once the combination of chosen strategies
$(s_A,s_B)$ is known, i.e., the utility functions mapping
$(s_A,s_B)$ on a number \cite{Mye91}. The expected
pay-off quantifies the preference of the players.

%

In these lecture notes the idea of identifying 
strategic moves with
quantum operations as introduced in Refs. 
\cite{Eisert,Meyer} is further developed. 
This approach
appears to be fruitful in at least two ways
\cite{Eisert,Meyer,Nature,Vaidman}.
On the one hand several recently proposed
applications of quantum information theory
can already be conceived as competitive 
situations where several parties 
with more or less opposed
motives interact. These parties may, for example,
apply quantum operations on a bi-partite
quantum system \cite{Plenio}. 
In the same context, 
quantum cloning has been
formulated as a game between two players \cite{Wer98}. 
Similarly, eavesdropping in quantum cryptography \cite{BB84}
can be regarded as a game between the eavesdropper
and the sender, and there are similarities
of the extended form of quantum versions of games
and quantum algorithms \cite{Shor,Ekert}.
On the other hand a generalisation of
the theory of decisions into the domain of
quantum probabilities seems interesting,
as the roots of game theory are partly in 
probability theory. In this context it is
of interest to investigate what solutions
are attainable if superpositions
of strategies are allowed for\cite{Eisert,Meyer}.

Game theory does not explicitly concern itself with 
how the information
is transmitted 
once a decision is taken.
Yet, it should
be clear that the practical implementation 
of any (classical) game inevitably makes use of the
the exchange of voting papers, faxes, emails, 
ballots, and the like.
In the Prisoners' Dilemma, e.g., the two parties
have to communicate with an advocate by talking to 
her or by writing a short letter on which
the decision is indicated.
%
%
%
%
Bearing in mind that a game
is also about the transfer of information, it
becomes legitimate to ask what happens if
these carriers of information are taken to be
quantum systems, quantum information being a 
fundamental notion of information. 

By classical means a two player binary choice
game may be played as follows:
An arbiter takes two 
coins 
and forwards one coin each to the players.
The players then receive their coin
with head up and 
may keep it as it is (''cooperate'')
or turn it upside down so that
tails is up (''defection''). 
Both players then return the coins to the
arbiter who calculates the players'
final pay-off corresponding to the combination
of strategies he obtains from the players.
Here, the coins serve as the physical
carrier of information in the game. 
In a quantum version of such a game quantum systems 
would be used as such carriers of information.
For a 
binary choice two player game
an implementation making use of minimal 
resources involves two qubits as physical
carriers. 

\section{Quantum Games and Quantum Strategies}

Any quantum system which can be manipulated
by two parties or more and
where the utility
of the moves can be reasonably
quantified, may be conceived as a quantum game.

%
%
A {\it two-player
quantum game}\/ $\Gamma=({\cal H},\rho,S_A,S_B,P_A,P_B)$
is completely specified by the underlying Hilbert space 
${\cal H}$ of the physical system, 
the initial state $\rho\in {\cal S}({\cal H})$,
where ${\cal S}({\cal H})$ is the 
associated state space,
the sets $S_A$ and $S_B$ of 
permissible quantum operations of
the two players, and
the {\it utility functionals}\/ 
$P_A$ and $P_B$, which specify the
utility for each player.
A {\it quantum strategy}\/ $s_A\in S_A$, $s_B\in S_B$
is a  quantum operation, that is,
a completely positive trace-preserving map mapping
the state space on itself \cite{Oper}.
The quantum game's definition also includes 
certain implicit rules, such as the order
of the implementation of the respective
quantum strategies.
Rules also exclude certain 
actions, as the alteration of the pay-off during the game.

The quantum games
proposed in Refs.\ \cite{Eisert}, \cite{Meyer}, and
\cite{Vaidman}
can be cast into this
form. Also, the
quantum cloning device as described in \cite{Wer98}
can be said to be a quantum game in this sense.
A quantum game is called 
{\it zero-sum game}\/, if the
expected pay-offs sum up to zero for all pairs of
strategies, that is, if 
$P_A(s_A,s_B)=-P_B(s_A,s_B)$
for all $s_A\in S_A$, $s_B\in S_B$. Otherwise, it is 
called a {\it non-zero sum game}\/.

%

It is natural to call two quantum strategies of 
Alice $s_A$ and
$s_A'$ equivalent, if $P_A(s_A,s_B)=P_A(s_A',s_B)$
and $P_B(s_A,s_B)=P_A(s_A',s_B)$ for {\it all}\/
possible $s_B$. That is, if 
$s_A$ and $s_A'$ yield the same expected
pay-off for both players
for all allowed strategies of Bob.
In the same way strategies $s_B$ and $s_B'$ 
of Bob will be identified.

\begin{figure}
\centerline{
	\epsfxsize=7.0cm
      \epsfbox{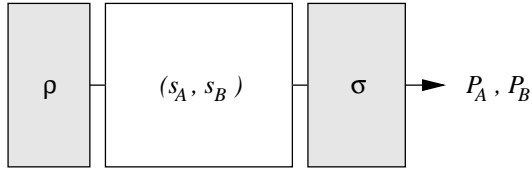}
}
\medskip

\caption{The general setup of a quantum game.}
\label{fig:Games}
\end{figure}

A solution concept provides advice to the players 
with respect to the
action they should take. 
The following solution concepts  will be  
used in the remainder of this lecture. These 
definitions are fully analogous to the
corresponding definitions in standard game
theory \cite{Mye91}.

A quantum strategy  $s_A$ 
is called {\it dominant strategy}\/ 
of Alice if
\begin{equation}
        P_A(s_A,s_B')
        \geq
        P_A(s_A',s_B')
\end{equation}
for all $s_A'\in S_A$, $s_B'\in S_B$.
Analogously we can define a dominant strategy
for Bob. 
A pair $(s_A,s_B)$ is said to be an {\it equilibrium 
in dominant strategies}\/ if $s_A$ and $s_B$ are
the players' respective dominant strategies.
A combination of strategies $(s_A,s_B)$
is called a {\it Nash equilibrium}\/ if
\begin{mathletters}
\begin{eqnarray}
        P_A(s_A,s_B)&&\geq P_A(s_A',s_B),\\
        P_B(s_A,s_B)&&\geq P_B(s_A,s_B')
\end{eqnarray}
\end{mathletters}
for all $s_A'\in S_A$, $s_B'\in S_B$.
A pair of strategies $(s_A, s_B)$ is called 
{\it Pareto optimal}\/, if it is not
possible to increase one player's pay-off
without lessening the pay-off of the other
player.

A solution in dominant strategies is the
strongest solution concept for a non-zero sum game. 
In the Prisoner's Dilemma defection is the 
dominant strategy, as it is favourable
regardless what strategy the other party
picks.
%
Typically, however,
the optimal strategy depends on the strategy chosen by the
other party. A Nash equilibrium implies that 
neither player has a motivation to unilaterally
alter his or her strategy from this equilibrium solution, 
as this action will lessen his or her pay-off.
Given that the other player will stick to the strategy
corresponding to the equilibrium, the best
result is achieved by also playing the equilibrium solution.
The concept of Nash equilibria is of paramount
importance to studies of non-zero-sum games.
It is, however, only an acceptable solution concept
if the Nash equilibrium is unique. 
For games with multiple equilibria 
the application of a 
hierarchy of natural refinement concepts may finally
eliminate all but one of the Nash equilibria.
Note that a Nash equilibrium is not necessarily 
efficient. In the Prisoners' Dilemma for example
there is a unique equilibrium, but it is not
Pareto optimal, 
meaning that there is another outcome
which would make both players better off.

\section{Two-qubit quantum games}

In the subsequent investigation we turn to
specific games where the classical
version of the game is faithfully
entailed in the quantum game.
%
In a quantum version of a 
binary choice game
two qubits are prepared by a arbiter in a 
particular initial state, the qubits are sent to
the two players who have physical instruments at
hand to manipulate their qubits in an appropriate
manner. Finally, the qubits are sent back
to the arbiter who performs a measurement to evaluate
the pay-off. 
%
For such a bi-partite quantum game
the system of interest 
is a quantum system with underlying Hilbert space
$
	{\cal H}={\cal H}_A\otimes{\cal H}_B$, 
${\cal H}_A={\cal H}_B={\Bbb{C}}^2$,
and associated 
state space ${\cal S}({\cal H})$.
Quantum strategies $s_A$ and $s_B$ of Alice and Bob
are local quantum operations
acting in ${\cal H}_A$ and ${\cal H}_B$ respectively
\cite{Ident}.
That is,
Alice and Bob are restricted to 
implement their respective 
quantum strategy
$s_A$ and $s_B$ on their qubit only.
In this step they may 
choose any quantum
strategy that is included
in the set of strategies $S$. They are
both well aware of the set $S$, but they do not
know which particular quantum strategy the
other party would actually implement. 
As the application of both quantum 
strategies amounts to a map
$
	s_A\otimes s_B:{\cal S}({\cal H})\rightarrow 
{\cal S}({\cal H})$,
after execution of the moves the system is 
in the state
\begin{equation}
	\sigma=(s_A\otimes s_B)(\rho).
\end{equation}
Particularly important will be unitary
operations $s_A$ and $s_B$. They are associated
with unitary operators $U_A$ and $U_B$,
written as $s_A\sim U_A$ and $s_B\sim U_B$.
In this case  
the final state $\sigma$ is  given by
\begin{equation}
	\sigma=(U_A\otimes U_B) \rho (U_A\otimes U_B)^\dagger.
\end{equation}
From now on both the sets of strategies of
Alice and Bob and
the pay-off functionals 
are taken to
be identical, that is,
\begin{equation}
S_A=S_B=S\,\,\,\,\text{ and }\,\,\,\,P_A=P_B=P,
\end{equation}
such 
that both parties face the same situation.

The quantum game $\Gamma=({\Bbb{C}}^2\otimes {\Bbb{C}}^2, \rho,
S,S,P,P)$ 
can be played in the following way:
The initial state $\rho$ is taken to be a maximally
entangled state in the respective state space.
In order to be consistent with Ref.\cite{Eisert}
let $\rho=|\psi\rangle\langle\psi|$ with
\begin{equation}\label{psis}
	|\psi\rangle=
		(|00\rangle+i |11\rangle)/\sqrt{2},
\end{equation}
where the first entry refers to ${\cal H}_A$ 
and the second to ${\cal H}_B$.
%
%
The two qubits are forwarded to the arbiter who performs
a projective selective measurement
on the final state $\sigma$
with Kraus operators
$\pi_{CC}$, $\pi_{CD}$, $\pi_{DC}$, and $\pi_{DD}$,
where
\begin{mathletters}
\begin{eqnarray}
 	\pi_{CC}&=&|\psi_{CC}\rangle\langle\psi_{CC}|,\,\,\,\,\,
	|\psi_{CC}\rangle=(|00\rangle+i |11\rangle)/\sqrt{2},\\
		\pi_{CD}&=&|\psi_{CD}\rangle\langle\psi_{CD}|,\,\,\,\,\,
	|\psi_{CD}\rangle=(|01\rangle-i |10\rangle)/\sqrt{2},
		\label{psi12}\\
		\pi_{DC}&=&|\psi_{DC}\rangle\langle\psi_{DC}|,\,\,\,\,\,
	|\psi_{DC}\rangle=(|10\rangle-i |01\rangle)/\sqrt{2},\\
	\pi_{DD}&=&|\psi_{DD}\rangle\langle\psi_{DD}|,\,\,\,\,\,
	|\psi_{DD}\rangle=(|11\rangle+i |00\rangle)/\sqrt{2}.
\end{eqnarray}
According to the outcome of the measurement, 
a pay-off of $A_{CC}$, $A_{CD}$, $A_{DC}$, or $A_{DD}$
is given to Alice, Bob receives 
$B_{CC}$, $B_{CD}$, $B_{DC}$, or $B_{DD}$.
The utility functionals, also referred to as
expected pay-off of Alice and Bob, read
\end{mathletters}
\begin{mathletters}
\begin{eqnarray}
	P_A(s_A,s_B)&=&A_{CC} {\rm tr}[\pi_{CC}
	\sigma]+
	A_{CD} {\rm tr}[\pi_{CD}
	\sigma]+
	A_{DC} {\rm tr}[\pi_{DC}
	\sigma]+
	A_{DD} {\rm tr}[\pi_{DD}
	\sigma],\label{PA}\\
	P_B(s_A,s_B)&=&
	B_{CC} {\rm tr}[\pi_{CC}
	\sigma]+
	B_{CD} {\rm tr}[\pi_{CD}
	\sigma]+
	B_{DC} {\rm tr}[\pi_{DC}
	\sigma]+
	B_{DD} {\rm tr}[\pi_{DD}
	\sigma].\label{PB}
\end{eqnarray}
It is important to note that the
Kraus operators are chosen 
in such a way that the classical game is fully entailed
in the quantum game: The
{\it classical strategies} $C$ and $D$ are associated
with particular unitary operations,
\begin{equation}
	C\sim 
	\left(
	\begin{array}{cc}
	1 & 0 \\
	0 & 1 
	\end{array}
	\right),\,\,\,\,\,
	D
	\sim
	\left(
	\begin{array}{cc}
	0 & 1 \\
	-1 & 0 
	\end{array}
	\right).
\end{equation}
$C$ does not change the state at all, $D$ implements a 
''spin-flip''. If both parties stick to these
classical strategies, Eq.\ (\ref{PA}) and Eq.\ (\ref{PB}) 
guarantee that 
the expected pay-off
is exactly the pay-off of the corresponding classical game
defined by the numbers $A_{CC}$, $A_{CD}$, $A_{DC}$, $A_{DD}$,
$B_{CC}$, $B_{CD}$, $B_{DC}$, and $B_{DD}$.
E.g., if Alice plays $C$ and Bob chooses $D$,
the state $\sigma$ after implementation of the
strategies is given by
\begin{equation}
	\sigma= (C\otimes D)(\rho)= 
	|\psi_{CD}\rangle
	\langle \psi_{CD}|,
\end{equation}
such that Alice obtains $A_{CD}$ units 
and Bob $B_{CD}$ 
units pay-off (see Fig.\ 1).
In this way
the peculiarities of 
strategic moves in the quantum domain
can be adequately studied. 
The players may make use of additional
degrees of freedom which are not available
with randomisation of the classical
strategies, but
they can also stick to mere
classical strategies.
This scheme can be applied to any
two player binary choice game and is
to a high extent canonical.\\

\end{mathletters}

\subsection{Prisoners' Dilemma}
%
We now investigate the 
solution concepts for the quantum analogue
of the 
Prisoners' Dilemma
(see Fig.\ \ref{Prison}) \cite{Values},
\begin{mathletters}
\begin{eqnarray}
	&&A_{CC}=B_{CC}=3,\,\,\,
	A_{DD}=B_{DD}=1,\\
	&&A_{CD}=B_{DC}=0,\,\,\,
	A_{DC}=B_{CD}=5.
\end{eqnarray}
In all of the following sets of 
allowed strategies $S$  the classical 
options (to defect and to cooperate)
are included. 
Several interesting sets of strategies
and concomitant solution concepts
will at this point be studied. The first three
subsections involve local unitary operations only,
while in the last subsection other quantum
operations are considered as well. 
\end{mathletters}

\subsubsection{ One-parameter set of strategies.}
The first set of strategies $S^{(CL)}$
involves 
quantum operations $s_A$ and $s_B$ 
which are local
rotations with one parameter.
The matrix representation of the
corresponding unitary operators
is taken to be
\begin{equation}
        {U}(\theta)=
        \left(
        \begin{array}{cc}
        \cos(\theta/2)&
        \sin(\theta/2)\\
        -\sin(\theta/2)
        &
		  \cos(\theta/2)
        \\
        \end{array}
        \right)
\end{equation}
with $\theta\in[0,\pi]$. Hence, 
in this simple case, selecting
strategies $s_A$ and $s_B$ amounts
to choosing two angles
$\theta_A$ and $\theta_B$. The 
classical strategies of defection and
cooperation are also included in the set of
possible strategies,
	$C\sim 
	U(0)$,
	$D
	\sim
	U(\pi)$.
An analysis of the expected pay-offs $P_A$ and $P_B$,
\begin{mathletters}
\begin{eqnarray}
	P_A(\theta_A,\theta_B)&=&
	3 |\cos(\theta_A/2)\cos(\theta_B/2)|^2
	+
	5 |\cos(\theta_B/2)\sin(\theta_A/2)|^2\nonumber\\
	&+&
	|\sin(\theta_A/2)\sin(\theta_B/2)|^2,\\
	P_B(\theta_A,\theta_B)&=&
	3 |\cos(\theta_A/2)\cos(\theta_B/2)|^2
	+
	5 |\sin(\theta_B/2)\cos(\theta_A/2)|^2\nonumber\\
	&+&
	|\sin(\theta_A/2)\sin(\theta_B/2)|^2,
\end{eqnarray}
shows that this game is the classical Prisoners'
Dilemma game \cite{Eisert}. 
The pay-off functions are actually identical
to the analogous functions in the ordinary Prisoners'
Dilemma with mixed (randomised)
strategies, 
where cooperation is chosen with the classical
probability 
$p=\cos^2(\theta/2)$. 
The inequalities
\end{mathletters}
\begin{mathletters}
\begin{eqnarray}
P_A(D,s_B)&\geq& P_A(s_A,s_B),\\
P_B(s_A,D)&\geq& P_B(s_A,s_B)
\end{eqnarray} 
hold
for all $s_A, s_B\in S^{(CL)}$, therefore,
$(D,D)$
is an equilibrium in dominant strategies and 
thus the unique Nash equilibrium. As explained in the 
introduction
this equilibrium is far from being efficient, 
because $P_A(D,D)=P_B(D,D)=1$
instead of the Pareto optimal pay-off which would
be 3.
\end{mathletters}

\subsubsection{Two-parameter set of strategies}
A more general set of strategies is
the following two-parameter set $S^{(TP)}$.
The matrix representation of 
operators corresponding to
quantum strategies from this set
is given by
\begin{equation}
        {U}(\theta,\phi)=
        \left(
        \begin{array}{cc}
        e^{i\phi}\cos(\theta/2)&
        \sin(\theta/2)\\
        -\sin(\theta/2)
        &e^{-i\phi}\cos(\theta/2)
        \\
        \end{array}
        \right)
\end{equation}
with $\theta\in[0,\pi]$ and $\phi\in[0,\pi/2]$. 
Selecting a strategy $s_A,s_B$
then means choosing
appropriate angles $\theta_A,\phi_A$ and $\theta_B,\phi_B$. 
The classical pure strategies can
be realised as 
\begin{equation}
C\sim U(0,0)\,\,\,\, \text{ and }\,\,\,\,
D\sim U(\pi,0).
\end{equation}
This case has also been considered in Ref. \cite{Eisert}.
The expected pay-off for Alice, e.g., explicitly
reads 
\begin{eqnarray}\label{LongPay}
	&&P_A(\theta_A,\phi_A,\theta_B,\theta_B)
	=3\left| \cos(\phi_A+\phi_B)\cos(\theta_A/2)\cos(\theta_B/2)\right|^2
	\\
		&+& 5
		\left|
			\sin(\phi_A)\cos(\theta_A/2)\sin(\theta_B/2)\right.
			-
			\left.
			\cos(\phi_B) \cos(\theta_B/2)\sin(\theta_A/2)
		\right|^2\nonumber\\
		&+& \left|
			\sin(\phi_A+\phi_B)\cos(\theta_A/2)\cos(\theta_B/2)
			\right.
			+
			\left.
			\sin(\theta_A/2)\sin(\theta_B/2)
		\right|^2.\nonumber
\end{eqnarray}
It turns out that the previous Nash equilibrium
$(D,D)$ of $S^{(CL)}$
is no longer an equilibrium solution, as
both players can benefit from deviating from
$D$. However, concomitant with the disappearance
of this solution another Nash equilibrium
has emerged, given
by $(Q,Q)$.  
The strategy $Q$ is associated with
a matrix
\begin{equation}
	Q\sim U(0,\pi/2)=
   \left(\begin{array}{cc} i & 0 \\ 0 &  
	-i\end{array}\right)\, .
\end{equation}
This Nash equilibrium is unique \cite{Eisert} and serves
as the only acceptable solution of the game.
The astonishing fact is that $P_A(Q,Q)=P_B(Q,Q)=3$
(instead of $1$)
so that the Pareto optimum is realised. 
No player could
gain without lessening the other player's expected pay-off.
In this sense one can say that the dilemma of
the original game has fully disappeared.
In the classical game only 
mutual cooperation is
Pareto optimal, but this pair of
strategies does not correspond
to a Nash equilibrium.

\subsubsection{General unitary operations}
One can generalise the previous setting to the
case where Alice and Bob can
implement operations $s_A$ and $s_B$ taken
from $S^{(GU)}$, where $S^{(GU)}$ is the set 
of general local 
unitary operations.
%
Here, it could be suspected that 
the solution becomes
more efficient the larger the
sets of allowed operations are. 
But this is not the case.
The previous Pareto optimal
unique Nash equilibrium $(Q,Q)$ ceases to be
an equilibrium solution if the set is enlarged: 
For any
strategy $s_B\in S^{(GU)}$
there exists an {\it optimal answer }
$s_A\in S^{(GU)}$ 
resulting in
\begin{equation}\label{Well}
	(s_A\otimes s_B)(\rho)=
	|\psi_{DC}\rangle\langle \psi_{DC}|,
\end{equation}
with $\rho$ given in Eq.\ (\ref{psis}).
%
That is, for any strategy of Bob $s_B$ 
there is a strategy $s_A$ of Alice such that
\begin{equation}
	P_A(s_A,s_B)=5\,\, \text{ and }\,\,
	P_B(s_A,s_B)=0:
\end{equation}
Take
\begin{eqnarray}
		s_A\sim
		\left(
		\begin{array}{cc}
		a & b\\
		c & d 
		\end{array}
		\right),\,\,\,
		s_B\sim 
		\left(
		\begin{array}{cc}
		-i b & a\\
		-d & -ic 
		\end{array}
		\right),
	\end{eqnarray}
	where $a,b,c,d$ are appropriate complex numbers.
Given that Bob plays the strategy $s_B$ associated
with a particular Nash equilibrium $(s_A,s_B)$, Alice
can always apply the optimal answer $s_A$ to achieve
the maximal possible pay-off. However, the resulting
pair of quantum strategies can not be an equilibrium
since again, the game being symmetric,
Bob can improve his pay-off by
changing his strategy to his optimal answer $s_B'$.
Hence, there is no pair $(s_A,s_B)$
of pure strategies with the property that the
players can only lose from unilaterally
deviating from this pair of strategies. 

Yet, there remain to be Nash equilibria in {\it mixed
strategies}\/ 
which are much more efficient than the classical 
outcome of the equilibrium in dominant strategies 
$P_A(D,D)=P_B(D,D)=1$. 
In a mixed strategy of Alice, say,
she selects a particular
quantum strategy $s_A$ (which is also called 
{\it pure strategy}\/)
from the set of strategies $S_A$
with a certain classical probability.
%
%
%
%
That is,
mixed strategies of Alice and Bob
are associated with
maps of the form 
\begin{equation}\label{doubly}
	\rho\longmapsto \sigma=\sum_{i,j} p^{(i)}_A p^{(j)}_B
	(U_A^{(i)}\otimes U_B^{(j)}) 
	\rho  
	(U_A^{(i)}\otimes U_B^{(j)})^\dagger, 
\end{equation}
$p^{(i)}_A, p^{(i)}_B \in [0,1]$, $i,j=1,2,...,N$, with
\begin{equation}
\sum_i p^{(i)}_A =\sum_j p^{(j)}_B =1.
\end{equation}
$U_A^{(i)}$
and $U_B^{(j)}$ are local unitary operators
corresponding to pure strategies $s_A^{(i)}$
and $s_B^{(j)}$. 

The map given by 
Eq.\ (\ref{doubly}) acts in ${\cal H}_A$ 
and  ${\cal H}_A$ 
as a doubly stochastic map, that is, as
a completely positive unital map \cite{Majo2}. 
As a result, 
the final reduced states 
${\rm tr}_B[\sigma]$ and 
${\rm tr}_A[\sigma]$
must be more mixed than the
reduced initial states 
${\rm tr}_B[\rho]$ 
and ${\rm tr}_A[\rho]$ 
in the sense of 
majorisation theory \cite{Majo}. 
As the initial state $\rho$ is a maximally entangled state,
all accessible states after application of a mixed strategy
of Alice and Bob are locally identical to the maximally mixed
state $1/{\rm dim}({\cal H}_A)=
1/{\rm dim}({\cal H}_B)$, which is a multiple
of $1$.

The following construction, e.g., yields an 
equilibrium in mixed quantum strategies:
Allow Alice to choose from two strategies
$s_A^{(1)}$ and $s_A^{(2)}$ with probabilities
$p^{(1)}_A=1/2$ and $p^{(2)}_A=1/2$,
while Bob may
take
$s_B^{(1)}$ or $s_B^{(2)}$, with
\begin{mathletters}
\begin{eqnarray}\label{PartNash}
	s_A^{(1)}\sim 
	\left(
	\begin{array}{cc}
        1 & 0 \\
	0 & 1 
        \end{array}
        \right),\,\,\,&&
	s_A^{(2)}\sim 
	\left(\begin{array}{cc}
        -i & 0 \\
	0 & i 
        \end{array}
        \right),\\
	s_B^{(1)}\sim 
	\left(\begin{array}{cc}
        0 & 1 \\
	-1 & 0 
        \end{array}
        \right),\,\,\,&&
	s_B^{(2)}\sim 
	\left(\begin{array}{cc}
        0 & -i \\
	-i & 0 
        \end{array}
        \right).\label{PartNash2}
\end{eqnarray}
His probabilities
are also given by $p^{(1)}_B=1/2$ and $p^{(2)}_B=1/2$.
The quantum strategies  of Eq.\ (\ref{PartNash}) and Eq.\ (\ref{PartNash2})
are mutually optimal answers
and have the property that 
\end{mathletters}
\begin{mathletters}
\begin{eqnarray}\label{satisfaction}
	P_A(s_A^{(i)},s_B^{(i)})=0,\,\,\,\,
	&&P_B(s_A^{(i)},s_B^{(i)})=5,\\
	P_A(s_A^{(i)},s_B^{(3-i)})=5,\,\,\,\,
	&&P_B(s_A^{(i)},s_B^{(3-i)})=0,\label{satisfaction2}
\end{eqnarray}
for $i=1,2$. 
\end{mathletters}
Due to the particular constraints
of Eq.\ 
(\ref{satisfaction}) and Eq.\ 
(\ref{satisfaction2}) there exists no other mixed strategy
for Bob yielding a better pay-off than the above mixed 
strategy, given that Alice sticks to the equilibrium
strategy.
This can be seen as follows. Let Alice use this particular
mixed quantum strategy as above and let Bob use
any mixed quantum strategy 
\begin{equation}\label{dono}
s_B^{(1)},..., s_B^{(N)}
\end{equation}
together with
 $p_A^{(1)},..., p_A^{(N)}$. The final state  $\sigma$
after application of the strategies
is given by the convex combination 
\begin{equation}\label{doubly}
	\sigma=\sum_{i=1,2}\sum_j p^{(i)}_A p^{(j)}_B
	(s_A^{(i)}\otimes s_B^{(j)})
	(\rho),
\end{equation}
This convex combination can only
lead to a smaller expected pay-off for Bob than 
the optimal pure strategy $s_B^{(k)}$ in Eq.\ (\ref{dono}), 
$k\in \{1,...,N\}$. 
Such optimal pure strategies
are given by $s_B^{(1)}$ and $s_B^{(2)}$ as in Eq.\ (\ref{PartNash2})
leading to an expected pay-off for Bob of $P_B(s_A,s_B)=2.5$;
there are no pure strategies which achieve a larger
expected pay-off. While both pure strategies $s_B^{(1)}$ and $s_B^{(2)}$
do not correspond to an equilibrium, the mixed strategy where
$s_B^{(1)}$ and $s_B^{(2)}$ are chosen with $p_B^{(1)}=1/2$
and $p_B^{(2)}=1/2$ actually does. Nash equilibria consist
of pairs of mutually optimal answers, and only for this choice
of Bob the original mixed quantum strategy of Alice is
{\it her}\/ optimal choice, as the same argument applies
also to her, the game being symmetric.

This Nash equilibrium is however not the only one. 
There exist also other four-tuples
of matrices than the ones presented
in  Eq.\ (\ref{PartNash}) and  Eq.\ (\ref{PartNash2}) 
that satisfy Eq.\ (\ref{satisfaction})
and  Eq.\ (\ref{satisfaction2}). 
Such matrices can be made out by appropriately
rotating the matrices of Eq.\ (\ref{PartNash}) and  
Eq.\ (\ref{PartNash2}).
In the light of the fact that there is more than
one equilibrium
it is not obvious which Nash equilibrium the players will 
realise. 
It is at first
not even 
evident whether a Nash equilibrium will be played at
all. But the game theoretical
concept of the {\it focal point effect}\/
\cite{Schelling,Mye91}
helps to resolve this issue. 

To explore the general structure of any Nash equilibrium
in mixed strategies we continue as follows: let  
\begin{equation}\label{Mono}
U_A^{(1)},..., U_A^{(N)}
\end{equation} 
together with
 $p_A^{(1)},..., p_A^{(N)}$ specify the mixed strategy
pertinent to a Nash equilibrium
of Alice. Then there exists a mixed strategy
$U_B^{(1)},..., U_B^{(N)}$, $p_B^{(1)},..., p_B^{(N)}$
of Bob which rewards Bob with the best achievable 
pay-off, 
given that Alice plays this mixed strategy.
Yet, the pair of mixed strategies associated with
\begin{equation}\label{Dual}
	Q U_A^{(1)} Q^\dagger ,..., Q U_A^{(N)} Q^\dagger,
	\,\,\,\,\,\,
	Q U_B^{(1)} Q^\dagger ,..., Q U_B^{(N)} Q^\dagger
\end{equation}
with 
$p_A^{(1)},..., p_A^{(N)}$, 
$p_B^{(1)},..., p_B^{(N)}$ is another Nash
equilibrium. This equilibrium leads to the same
expected pay-off for both players, and is fully
symmetric to the previous one.
Doubly applying $Q$ as
$QQU_A^{(1)} Q^\dagger
Q^\dagger ,..., Q Q U_A^{(N)} Q^\dagger Q^\dagger$
results again into a situation with equivalent
strategies as the original ones.
For a given Nash equilibrium as above the one
specified by Eq.\ (\ref{Dual})
will be called dual equilibrium.

However, there is a single Nash equilibrium $(R,R)$
which is the only one which gives an expected pay-off
of $P_A(R,R)=P_B(R,R)=2.25$ and which is identical
to its dual equilibrium: it is the simple map
\begin{equation}\label{SimpleMap}
	\rho\longmapsto \sigma=1/{\rm dim}({\cal H}).
\end{equation}
Indeed, there exist probabilities
$p_A^{(1)},...,p_A^{(N)}$
and unitary operators $U_A^{(1)},...,U_A^{(N)}$
such that $\sum_i p_A^{(i)}
	(U_A^{(i)}\otimes 1) \rho  (U_A^{(i)}
	\otimes 1)^\dagger=1/{\rm dim}({\cal H})$
\cite{Majo}.
If Alice has already selected $s_A=R$, the application
of $s_B=R$ will not change the state of the 
quantum system any more.

Assume that Eq.\ (\ref{Mono}) and Eq.\ (\ref{Dual})
are associated with equivalent quantum strategies.
This means that they have to produce the same
expected pay-off for all quantum strategies $s_B$
of Bob. 
If Alice and Bob apply $s_A\otimes s_B$ they
get an expected pay-off according to 
Eq.\ (\ref{PA}) and Eq.\ (\ref{PB});
if Alice after implementation of $s_A$ manipulates
the quantum system by applying the local unitary
operator $Q\otimes 1$, they obtain 
\begin{mathletters}
\begin{eqnarray}
	P'_A(s_A,s_B)&=&A_{DD} {\rm tr}[\pi_{CC}
	\sigma]+
	A_{DC} {\rm tr}[\pi_{CD}
	\sigma]+
	A_{CD} {\rm tr}[\pi_{DC}
	\sigma]+
	A_{CC} {\rm tr}[\pi_{DD}
	\sigma],\\
	P'_B(s_A,s_B)&=&
	B_{DD} {\rm tr}[\pi_{CC}
	\sigma]+
	B_{DC} {\rm tr}[\pi_{CD}
	\sigma]+
	B_{CD} {\rm tr}[\pi_{DC}
	\sigma]+
	B_{CC} {\rm tr}[\pi_{DD}
	\sigma].
\end{eqnarray}
The only $s_A$ with the property that $P'_A(s_A,s_B)=P_A(s_A,s_B)$
and $P'_B(s_A,s_B)=P_B(s_A,s_B)$
for all $s_B$ is the map given by Eq.\ (\ref{SimpleMap}).
\end{mathletters}

In principle, any Nash equilibrium may become a self-fulfilling
prophecy if the particular Nash equilibrium is expected by
both players. It has
been pointed out that
in a game with more than one equilibrium, anything that 
attracts the players' attention towards one of the equilibria
may make them expect and therefore realise it
\cite{Schelling}. 
The corresponding {\it focal equilibrium}\/ \cite{Mye91}
is the one which is conspicuously
distinguished from the other Nash equilibria. 
In this particular situation there is indeed one Nash equilibrium
different from all the others: it is the one which is
equivalent to its dual equilibrium, the map which simply
maps the initial state on the maximally mixed state.
For all other expected pay-offs both players are ambivalent between
(at least) two symmetric equilibria. 
The expected pay-off the players will receive in this
focal equilibrium --
\begin{equation}
P_A(R,R)=P_B(R,R)=2.25
\end{equation} 
-- is not fully
Pareto optimal, but it is again much more efficient than the 
classically achievable outcome of 1 \cite{NewRemark}.

\subsubsection{
Completely positive trace-preserving maps corresponding to 
local operations}
In this scenario both Alice and Bob
may perform any operation that is allowed by quantum mechanics.
That is, the set of strategies $S^{(CP)}$
is made up of $(s_A, s_B)$, where
both $s_A$ and $s_B$ correspond to a completely
positive trace-preserving map 
\begin{equation}
	(s_A\otimes s_B) (\rho)= \sum_i \sum_j (A_i \otimes B_j)  
	\rho ( A_i\otimes B_j) ^\dagger 
\end{equation}
corresponding to a 
local operation, 
associated with Kraus operators $A_i$ and $B_j$ with 
$i,j=1,2,...$ .
The trace-preserving property requires 
$\sum_i A_i^\dagger A_i=1$ and $\sum_i B_i^\dagger B_i=1$.
This case has already been mentioned in Ref.\  \cite{Eisert}.
The quantum strategies $s_A$ and $s_B$ 
do no longer inevitably
act as unital maps in the
respective Hilbert spaces as before.
In other words, the reduced states
of Alice and Bob after application
of the quantum strategy are not 
necessarily identical to 
the maximally mixed state 
$1/{\rm dim}({\cal H}_A)$.

As already pointed out in Ref.\ \cite{Eisert},
the pair of strategies $(Q,Q)$ of the two-parameter 
set of strategies $S^{(TP)}$ is again no equilibrium solution.
It is straightforward to prove that the 
Nash equilibria of the type of Eq.\ (\ref{PartNash})
and Eq.\ (\ref{PartNash2})
of mixed strategies with general local 
unitary operations are, however, 
still present, and each
of these equilibria yields an expected pay-off of
$2.5$. 

In addition, as strategies do no longer have to 
be locally unital maps, it is not surprising
that new Nash equilibria emerge:
Alice and Bob may, e.g., perform 
a measurement associated with Kraus operators
\begin{equation}
	A_1=|0\rangle\langle0|\,\,\,\,\,A_2=
	|1\rangle\langle1|,\,\,\,\,\,
	B_1=D|0\rangle\langle0|\,\,\,\,\,B_2=
	D |1\rangle\langle1|.
\end{equation}
This operation yields a final state 
$\sigma=(s_A\otimes s_B)(\rho)=
(|01\rangle\langle01|+
|10\rangle\langle10|)/2$. 
Clearly neither Alice nor Bob can gain from
unilaterally deviating from their strategy. 

One can nevertheless argue as in the previous 
case. Again, all Nash equilibria occur at least
in pairs. First, there are again the 
dual equilibria from $S^{(GU)}$. Second,
there are Nash equilibria $(s_A,s_B)$, 
$s_A \neq s_B$,
with the property that $(s_B,s_A)$ is also 
a Nash equilibrium yielding the same expected
pay-off. The only Nash equilibrium invariant
under application of $Q$ 
and exchange of the strategies of the
players is again 
$(R,R)$ defined in the
previous subsection, which yields a pay-off
$P_A(R,R)=P_B(R,R)=2.25$.
%
This is the solution of the game is the
most general case. While
both players could in principle do better
(as the solution lacks Pareto optimality),
the efficiency of this focal equilibrium is
much higher than the equilibrium in 
dominant strategies of the classical game.
Hence, also in this most general case both players gain
from using quantum strategies.

This study shows that the 
efficiency of the equilibrium 
the players can reach in
this game depends on the actions the players may
take. One feature, however is present in
each of the considered sets:
both players can increase their expected
pay-offs drastically by resorting to quantum strategies.
\\

\subsection{Chicken}
\begin{figure}
\centerline{
        \epsfxsize=5.0cm
       \epsfbox{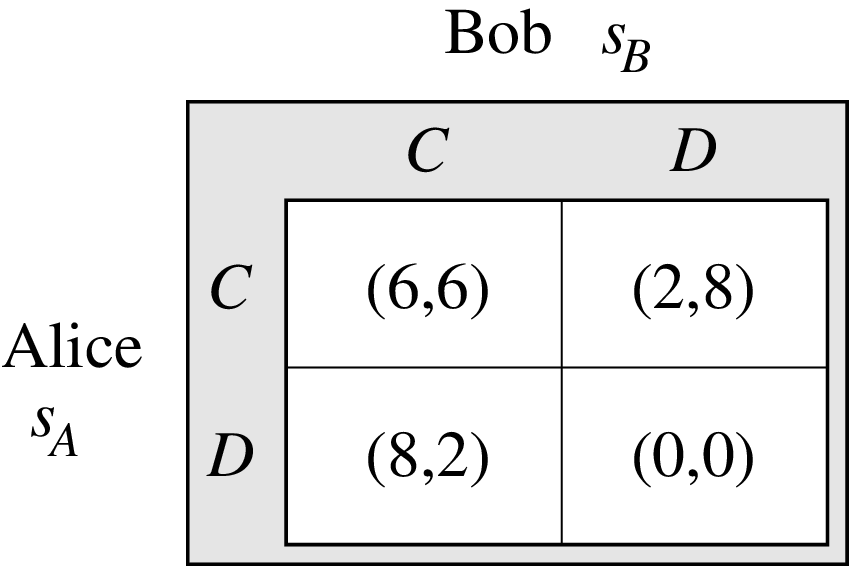}
}
\medskip
\caption{The pay-off matrix of the so-called Chicken game.}
\label{BattleFig}
\end{figure}

In the previous classical game -- the Prisoners' Dilemma -- 
an unambiguous solution can be specified consisting of
a unique Nash equilibrium. 
However, this solution was not
efficient, thus giving rise to the dilemma. The situation of the
players in the Chicken game \cite{Mye91,Pou92}, 
\begin{mathletters}
\begin{eqnarray}
        &&A_{CC}=B_{CC}=6,\,\,\,
        A_{CD}=B_{DC}=8,\\
        &&A_{DC}=B_{CD}=2,\,\,\,
        A_{DD}=B_{DD}=0,
\end{eqnarray}
can be described by the matrix of Fig.\ \ref{BattleFig}.
\end{mathletters}

\noindent 
This game
has two Nash
equilibria $(C,D)$ and $(D,C)$:
it is not clear how to anticipate
what the players' decision would be. In
addition to the
two Nash equilibria in pure strategies there is an
equilibrium in mixed strategies, yielding an expected
pay-off $4$ \cite{Mye91}.

In order to investigate the new features of the
game if superpositions of classical strategies 
are allowed for, three set of strategies are
briefly discussed:

\subsubsection{One-parameter set of strategies}
Again,
we consider the set of strategies $S^{(CL)}$ of
one-dimensional rotations. The strategies
$s_A$ and $s_B$ are associated with local
unitary operators
\begin{equation}
        {U}(\theta)=
        \left(
        \begin{array}{cc}
        \cos(\theta/2)&
        \sin(\theta/2)\\
        -\sin(\theta/2)
        &
                  \cos(\theta/2)
        \\
        \end{array}
        \right)
\end{equation}
with $\theta\in[0,\pi]$,
\begin{equation}
        C\sim 
        U(0)=\left(
        \begin{array}{cc}
        1 & 0 \\
        0 & 1 
        \end{array}
        \right),\,\,\,\,\,
        D
        \sim
        U(\pi)=\left(
        \begin{array}{cc}
        0 & 1 \\
        -1 & 0 
        \end{array}
        \right).
\end{equation}
Then as before, the 
quantum game yields the same expected pay-off
as the classical game in randomised strategies. 
This means that still two Nash equilibria in pure
strategies are present.

\subsubsection{Two-parameter set of strategies}
The players can actually take advantage of
an additional degree of freedom which is not accessible
in the classical game. If they may apply
unitary operations from  $S^{(TP)}$ of the type
\begin{equation}
        {U}(\theta,\phi)=
        \left(
        \begin{array}{cc}
        e^{i\phi}\cos(\theta/2)&
        \sin(\theta/2)\\
        -\sin(\theta/2)
        &e^{-i\phi}\cos(\theta/2)
        \\
        \end{array}
        \right)
\end{equation}
with $\theta\in[0,\pi]$ and $\phi\in[0,\pi/2]$
the situation is quite different than with $S^{(CL)}$.
$(C,D)$ and $(C,D)$ with
$C\sim U(0,0)$ and $D\sim U(\pi,0)$ are no
longer equilibrium solutions. E.g., given that
$s_A=D$ the pair of strategies
$(D,Q)$ with $Q\sim U(0,\pi/2)$ yields a better
expected pay-off for Bob than $(D,C)$, that is
to say $P_B(D,Q)=8$, $P_B(D,C)=2$. 
In fact
$(Q,Q)$ is now the unique Nash equilibrium with
$P_A(Q,Q)=P_B(Q,Q)=6$, which follows from an
investigation of the actual expected 
pay-offs of Alice and Bob analogous to Eq.\ (\ref{LongPay}).
This solution is not only the 
unique acceptable solution of the game, but
it is also an equilibrium that is Pareto optimal.
This contrasts very much with the situation
in the classical game, where the two
equilibria were not that efficient.

\subsubsection{Completely positive 
trace-preserving maps corresponding to 
local operations}
As in the considerations
concerning the Prisoner's Dilemma game,
more than one Nash equilibrium is present, 
if both players can take quantum strategies from the
set  $S^{(CP)}$,
and all Nash equilibria emerge at least in pairs
as above. The focal equilibrium is given by $(R,R)$,
resulting in a pay-off of 
$P_A(R,R)=P_B(R,R)=4$, which is the same
as the mixed strategy of the classical game.

\section{Summary and Conclusion}
In these lecture notes 
the idea of implementing
quantum operations as strategic moves in 
a game is explored \cite{Eisert}.
In detail, we investigated
games which could be conceived as a generalisation
into the quantum domain of a two player
binary choice game. As a toy model for
more complex scenarios we studied quantum
games where the efficiency of the equilibria 
attainable when using quantum strategies could be
contrasted with the efficiency of solutions in the
corresponding classical game. 
We investigated a hierarchy of 
quantum strategies
$S^{(CL)}
\subset S^{(TP)}
\subset S^{(GU)}
\subset S^{(CP)}$.
Again \cite{Eisert,Meyer}, we found superior
performance of quantum 
strategies as compared to classical strategies.

The nature of a game is determined by
the rules of the game. In particular,
the appropriate solution concept depends on
the available strategic moves. Obviously, a
player cannot make a meaningful choice without
knowing the options at his or her disposal.
So it comes to no surprise that also 
the actual achievable pay-off in such a 
game depends on the set of allowed strategies.
Roughly speaking, one can say that the possibility
of utilising strategies which are not feasible
in the analogous classical game implicates
a significant advantage. In the models studied
in detail two kinds of ''dilemmas'' were
''resolved'': (i)
On the one hand 
there are quantum games with an efficient 
unambiguous solution,
while in the classical analogue only an inefficient
equilibrium can be identified. By taking advantage
of appropriate quantum strategies much 
more efficient equilibria
could be reached. In certain sets of strategies
even a maximally efficient solution  -- the Pareto
optimum -- was attainable.
(ii) On the other hand, there exist quantum games with
a unique solution with a classical equivalent which
offers two Nash equilibria of the same quality.

This paper deals with simple set-ups in which 
information is exchanged quantum-mechanically.
The emphasis was to examine how situations
where strategies are
identified with quantum operations applied on
quantum mechanical carriers of information 
are different from the classical equivalent.
It is the hope that these investigations may
enable us to better understand
competitive structures 
in a game theoretical
sense in applications of quantum information
theory.

\section{Acknowledgements}
We would like to thank Maciej Lewenstein,
Onay Urfal{\i}o$\overline{\mathrm g}$lu,
Joel Sobel, 
Tom Cover, Charles H.\ Bennett, Martin B.\ Plenio,
and Uta Simon
for helpful suggestions. We also acknowledge fruitful 
discussions with the participants of the A2 Consortial 
Meeting. This work was supported by the European Union and 
the DFG.


\begin{references}
\bibitem{Oper}
	E.g., Alice may 
	allow a coupling of the original quantum system
	to an auxiliary quantum system and let the two unitarily
	interact. After performing a projective 
	measurement 
	on the composite system she could eventually 
	consider
	the original system again
	by taking a partial trace with respect
	to the auxiliary part.
\bibitem{Ident}
	 The quantum strategies		  
	$s_A\otimes 1$ and
	$1\otimes s_B$
	are in the following identified with $s_A$ and $s_B$,
	respectively.
\bibitem{NewRemark}
	In the classical game both players could also play
	$C$ and $D$ with probabilities $p=1/2$ yielding the
	same expected pay-off of $2.25$ for both players,
	but this pair of mixed strategies would be no equilibrium solution, 
	as any player could benefit from simply choosing the
	dominant strategy $D$.
\end{references}
\end{document}